\documentclass[10pt]{llncs}
\usepackage{placeins} 
\usepackage{underscore} 
\usepackage{lipsum}
\usepackage[font=itshape]{quoting} 
\usepackage[title]{appendix} 
\usepackage{graphicx} 
\usepackage{epstopdf} 
\usepackage[font=small,labelfont=bf]{caption}
\usepackage{textcomp} 
\usepackage{amsmath} 
\usepackage{pbox} 
\usepackage{hyperref} 
\usepackage{cite} 
\usepackage{url} 
\usepackage[misc]{ifsym} 

\newcommand*\justify{
  \fontdimen2\font=0.4em
  \fontdimen3\font=0.2em
  \fontdimen4\font=0.1em
  \fontdimen7\font=0.1em
  \hyphenchar\font=`\-
}
\usepackage{colortbl} 
\usepackage{array}
\usepackage{booktabs}
\usepackage{multirow}

\usepackage{amssymb}
\usepackage{pifont}
\newcommand{\cmark}{\ding{51}}%
\newcommand*\rot{\rotatebox{90}}
\usepackage{tabularx}
\usepackage{threeparttable, tablefootnote}

\usepackage{msc} 
\usepackage{xcolor}
\newcommand{\quotes}[1]{``#1''} 

\begin{document}
\title{What's in a Downgrade? A Taxonomy of Downgrade Attacks in the TLS Protocol and Application Protocols Using TLS}

\author{Eman Salem Alashwali\inst{1,}\inst{2}\textsuperscript{ (\Letter)} \and Kasper Rasmussen\inst{1}}  
\institute{University of Oxford, Oxford, United Kingdom \\
\email{\{eman.alashwali,kasper.rasmussen\}@cs.ox.ac.uk} \\
\and King Abdulaziz University (KAU), Jeddah, Saudi Arabia \\
\email{ealashwali@kau.edu.sa}
}
\maketitle
\setcounter{footnote}{0}

\begin{abstract}
A number of important real-world protocols including the Transport Layer Security (TLS) protocol have the ability to negotiate various security-related choices such as the protocol version and the cryptographic algorithms to be used in a particular session. Furthermore, some insecure application-layer protocols such as the Simple Mail Transfer Protocol (SMTP) negotiate the use of TLS itself on top of the application protocol to secure the communication channel. These protocols are often vulnerable to a class of attacks known as \emph{downgrade attacks} which targets this negotiation mechanism. In this paper we create the first taxonomy of TLS downgrade attacks. Our taxonomy classifies possible attacks with respect to four different vectors: the protocol element that is targeted, the type of vulnerability that enables the attack, the attack method, and the level of damage that the attack causes. We base our taxonomy on a thorough analysis of fifteen notable published attacks. Our taxonomy highlights clear and concrete aspects that many downgrade attacks have in common, and allows for a common language, classification, and comparison of downgrade attacks. We demonstrate the application of our taxonomy by classifying the surveyed attacks. 
\end{abstract}

\section{Introduction}\label{introduction}
A number of important real-world protocols, such as the Transport Layer Security protocol (TLS) \cite{tls12}\cite{tls13rev25}, which is used by billions of people everyday to secure internet communications, support multiple protocol versions and algorithms, and allow the communicating parties to negotiate them during the handshake. Furthermore, some important legacy application-layer protocols that are \textit{not} secure by design such as the Simple Mail Transfer Protocol (SMTP) \cite{smtp2001} allow the communicating parties to negotiate upgrading the communication channel to a secure channel over a TLS layer. However, experience has shown that protocol developers tend to maintain support for weak protocol versions and algorithms, mainly to provide backward compatibility. In addition, empirical analysis of real-world deployment shows that a high percentage of SMTP servers that support TLS and are capable of upgrading SMTP to SMTP-Secure (SMTPS) are configured in the \quotes{opportunistic security} mode \cite{opportunistic2014}, meaning that they \quotes{fail open}, and operate in an unauthenticated plaintext mode if the upgrade failed for any reason, favoring functionality over security \cite{bursztein15}\cite{durumeric15}. \par 

In a typical downgrade attack, an active network adversary\footnote{Throughout the paper we will use the terms: active network attacker, active network adversary, and man-in-the-middle interchangeably} interferes with the protocol messages, leading the communicating parties to operate in a mode that is weaker than they prefer and support. In recent years, several studies illustrated the practicality of downgrade attacks in widely used protocols such as TLS. More dangerously, downgrade attacks can succeed even when only one of the communicating parties supports weak choices as in \cite{adrian15}\cite{beurdouche15}. \par

There are plenty of reported downgrade attacks in the literature that pertain to TLS such as \cite{durumeric15}\cite{adrian15}\cite{beurdouche15}\cite{bhargavan2016transcript}\cite{moller14}\cite{ durumeric17}\cite{wagner96}\cite{bhargavan2016downgrade}\cite{Mavrogiannopoulos12}\cite{aviram16}. A close look at these attacks reveals that they are not all identical: they target various elements of the protocol, exploit various types of vulnerabilities, use various methods, and result in various levels of damage. \par 

The existing literature lacks a taxonomy that shows the big picture outlook of downgrade attacks that allows classifying and comparing them. To bridge this gap, this paper presents a taxonomy of downgrade attacks with a focus on the TLS protocol based on an analysis of fifteen notable published attacks. The taxonomy helps in deriving answers to the following questions that arise in any downgrade attack:

\begin{enumerate}
\item \textit{What has been downgraded?}
\item \textit{How is it downgraded?}
\item \textit{What is the impact of the downgrade?}
\end{enumerate}  

Our downgrade attack taxonomy classifies downgrade attacks with respect to four vectors: element (to answer: What has been downgraded?), vulnerability and method (to answer: How is it downgraded?), and damage (to answer: What is the impact of the downgrade?). \par 

The aim of this paper is to provide a reference for researchers, protocol designers, analysts, and developers that contributes to a better understanding of downgrade attacks and its anatomy. Although our focus in this paper is on the TLS protocol and the application protocols that use it, this does not limit the paper's benefit to TLS. The paper can benefit the design, analysis, and implementation of any protocol that has common aspects of TLS. \par   

Our contribution is twofold: First, we provide the first taxonomy of TLS downgrade attacks based on a thorough analysis of fifteen surveyed attacks. Our taxonomy dissects complex downgrade attacks into clear categories and provides a clean framework for reasoning about them. Second, although our paper is not meant to provide a comprehensive survey, however, as a necessary background, we provide a brief survey of all notable published TLS downgrade attacks. Unlike the existing general surveys on TLS attacks, our survey is focused on a particular family of attacks that are on the rise, and covers some important recent downgrade attacks that none of the existing surveys \cite{meyer13}\cite{clark13} (which date back to 2013) have covered. 

The rest of the paper is organised as follows: in section \ref{related-work}, we summarise related work. In section \ref{example}, we provide an illustrative example of downgrade attacks. In section \ref{model}, we describe the attacker model that we consider in our taxonomy. In section \ref{methodology}, we describe the methodology we use to devise the taxonomy. In section \ref{downgrade-attacks}, we briefly survey fifteen cases of downgrade attacks in TLS. In section \ref{taxonomy}, we present our taxonomy. In section \ref{discussion}, we provide a discussion. In section \ref{conclusion}, we conclude. Finally, Appendix \ref{appendix:a} provides a background in the TLS protocol.   

\section{Related Work} \label{related-work}
Bhargavan et al. \cite{bhargavan2016downgrade} provide a formal treatment of downgrade resilience in cryptographic protocols and define downgrade security. In our work, we look at downgrade attacks from an informal and pragmatic point of view. We also consider downgrade attacks in a context beyond the key-exchange, e.g. in negotiating the use of TLS layer in multi-layers protocols such as SMTP. \par

The work of \cite{clark13} and \cite{meyer13} provide surveys on TLS attacks in general. Their surveys cover some of the TLS downgrade attacks that we cover. However, our work is not meant to survey TLS downgrade attacks, but to analyse them to create a taxonomy of downgrade attacks and to provide a framework to reason about them. Furthermore, our work covers state-of-the-art TLS downgrade attacks that have not been covered in previous surveys such as downgrade attacks in draft-10 of the coming version of TLS (TLS~1.3) \cite{bhargavan2016downgrade}, the SLOTH attack \cite{bhargavan2016transcript}, the DROWN attack \cite{aviram16}, among others.\par 

Howard and Longstaff \cite{howard98} present a general taxonomy of computer and network attacks. Our approach is similar to the one taken in \cite{howard98} in terms of presenting the taxonomy in logically connected steps. We have some common categories such as the vulnerability, but we also introduce our own novel categories such as the element and damage which classifies downgrade attacks at a lower level. \par

In \cite{stricot16} a taxonomy of man-in-the-middle attacks is provided. It is based on four tiers: \quotes{state}, \quotes{target}, \quotes{behaviour}, and \quotes{vulnerability}. Our taxonomy is particularly focused on downgrade attacks, thus provides further insights over the general man-in-the-middle taxonomy. We also have different perspectives. For example, although we share the vulnerability category, \cite{stricot16} present it in an exhaustive list of vulnerabilities such as \quotes{cipher block chaining}, \quotes{compression}, \quotes{export key}, etc. while our approach is to focus on the source of the flaw that allows the attack. We end up with three vulnerability sub-categories: implementation, design, and trust-model, which are more likely to capture future attacks. 

\section{Downgrade Attacks, an Illustrative Example} \label{example}
\begin{figure*}[!tp]
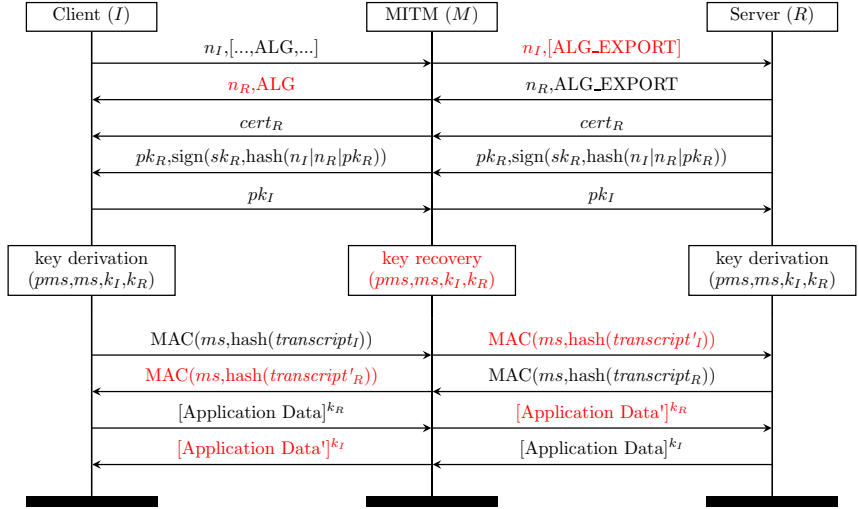
 
\centering
\resizebox{\textwidth}{!}{
\setmsckeyword{} 
\drawframe{no} 
\begin{msc}[normal values, /msc/level height=0.7cm, /msc/label distance=0.1cm , /msc/first level height=0.6cm, /msc/last level height=0.6cm]{}
\setlength{\instwidth}{2.5\mscunit} 
\setlength{\instdist}{4\mscunit} 
\declinst{I}{}{Client ($I$)}%
\declinst{M}{}{MITM ($M$)}%
\declinst{R}{}{Server ($R$)}%

\mess{$n_I$,[...,ALG,...]} {I}{M}
\mess{\textcolor{red}{$n_I$,[ALG_EXPORT]}}{M}{R} 
\nextlevel

\mess{$n_R$,ALG_EXPORT} {R}{M}
\mess{\textcolor{red}{$n_R$,ALG}}{M}{I}
\nextlevel

\mess{$cert_R$} {R}{M}
\mess{$cert_R$} {M}{I}
\nextlevel

\mess{$pk_{R}$,sign($sk_R$,hash($n_I|n_R|pk_{R}$))}{R}{M}
\mess{$pk_{R}$,sign($sk_R$,hash($n_I|n_R|pk_{R}$))}{M}{I}

\nextlevel
\mess{$pk_{I}$}{I}{M}
\mess{$pk_{I}$}{M}{R}

\nextlevel
\action*{\parbox{3cm} {\centering key derivation \\ ($pms$,$ms$,$k_I$,$k_R$)}}{I}
\action*{\parbox{3cm} {\centering \textcolor{red}{key recovery \\ ($pms$,$ms$,$k_I$,$k_R$)}}}{M}
\action*{\parbox{3cm} {\centering key derivation \\ ($pms$,$ms$,$k_I$,$k_R$)}}{R}
\nextlevel
\nextlevel
\nextlevel

\mess{MAC($ms$,hash(\textit{transcript\textsubscript{I}}))} {I}{M}
\mess{\textcolor{red}{MAC($ms$,hash(\textit{transcript\textquotesingle{}\textsubscript{I}}))}} {M}{R}

\nextlevel
\mess{MAC($ms$,hash(\textit{transcript\textsubscript{R}}))} {R}{M}
\mess{\textcolor{red}{MAC($ms$,hash(\textit{transcript\textquotesingle{}\textsubscript{R}}))}} {M}{I}

\nextlevel
\mess{[Application Data]\textsuperscript{$k_R$}} {I}{M}
\mess{\textcolor{red}{[Application Data\textquotesingle{}]\textsuperscript{$k_R$}}} {M}{R}

\nextlevel
\mess{[Application Data]\textsuperscript{$k_I$}} {R}{M}
\mess{\textcolor{red}{[Application Data\textquotesingle]\textsuperscript{$k_I$}}} {M}{I}
\end{msc}

} 
\caption{Illustrative example of downgrade attack in a simplified version of TLS.}
\label{fig:downgrade-illustration} 
\end{figure*}

Figure \ref{fig:downgrade-illustration} shows an illustrative example of downgrade attacks in a simplified version of the TLS~1.2 protocol inspired by the Logjam attack \cite{adrian15}. Throughout the paper in the message sequence diagrams, we denote the communicating parties by client (initiator $I$) and server (responder $R$). We denote the man-in-the-middle by (MITM $M$). A background on the TLS protocol that is necessary to comprehend the example is provided in Appendix \ref{appendix:a}.  

In this example, we assume certificate-based unilateral server-authentication mode using ephemeral Diffie-Hellman (DHE) key-exchange algorithm, and Message Authentication Code (MAC) to authenticate the exchanged handshake messages (the transcript). As depicted in Figure \ref{fig:downgrade-illustration} the client starts the handshake by sending its nonce ($n_I$) and a list of ciphersuites ([...,ALG,...]) to the server. The ciphersuite is a string (ALG) that defines the algorithms to be used in a particular session. In this example, we assume that the client's ciphersuites list contains only strong ciphersuites. The server must select one of the offered ciphersuites to be used in subsequent messages of the protocol. A man-in-the-middle modifies the client's proposed ciphersuites such that they offer only export-grade\footnote{Export-grade ciphers are weak ciphers with a maximum of 512-bit key for asymmetric encryption, and 40-bit key for symmetric encryption \cite{export}.} ciphersuite ([ALG_EXPORT]), e.g. key-exchange with 512-bit DHE group. If the server supports export-grade ciphersuites, for example, to provide backward compatibility to legacy clients, it will select an export-grade one, misguided by the modified client message that offered only export-grade ciphersuites. Then, the server sends its nonce ($n_R$) and its selected ciphersuite ALG_EXPORT to the client. To avoid detection, the man-in-the-middle modifies the server's choice from ALG_EXPORT to ALG to make it acceptable for the client that may not support export-grade ciphersuites as is the case in most updated web browsers today. Then, the server sends its certificate ($cert_R$), followed by a message that contains the server's public-key parameter $pk_R$, and a signed hash of the nonces ($n_I$ and $n_R$) and the server's public-key parameters $pk_R$. The signature is used to authenticate the nonces and the server's selected key parameters. However, in TLS~1.2 and below, the server's signature does not cover the server's selected ciphersuite (ALG_EXPORT in our example). Therefore, even if the client supports only strong ciphersuites, if it accepts arbitrary key parameters (e.g. non standard DHE groups), it will not distinguish whether the selected ciphersuite is export-grade or strong, and will generate weak keys based on the server's weak key parameters, despite the client's support for only strong ciphersuites. After that, the client sends its key parameter ($pk_I$). Then, both parties should be able to compute the pre-master secret ($pms$), the master secret ($ms$), and the client and server session keys, ($k_I$) and ($k_R$), respectively. The exchanged weak public-key parameters enable a man-in-the-middle to recover secret values from the weak public-keys, e.g. recover the private exponent from one or both parties' public-keys using Number Field Sieve (NFS) discrete log ($dlog$) (since we assume DHE key). Consequently, be able to compute the $pms$, $ms$, $k_I$, and $k_R$ in real-time. As a result of breaking the $ms$, the attacker can forge the MACs that are used to provide transcript integrity and authentication, hence, circumvent downgrade detection. Since the man-in-the-middle has the session keys, he can decrypt messages between the client and server as illustrated in Figure \ref{fig:downgrade-illustration}. This general example is similar to the Logjam \cite{adrian15} attack. This example is not the only form of TLS downgrade attacks as the paper will elaborate in the coming sections. 

\section{Attacker Model} \label{model}
In our taxonomy, we assume an external man-in-the-middle attacker who can passively eavesdrop on, as well as actively inject, modify, or drop messages between the communicating parties. The attacker can also connect to multiple servers in parallel. Furthermore, the attacker has access to bounded computational resources that allow him to break weak cryptographic primitives. 
 
\section{Methodology} \label{methodology}
First, to devise the taxonomy, we analyse fifteen published cases of downgrade attacks that relate to TLS from: \cite{durumeric15}\cite{adrian15}\cite{beurdouche15}\cite{bhargavan2016transcript}\cite{moller14}\cite{ durumeric17}\cite{wagner96}\cite{bhargavan2016downgrade}\cite{Mavrogiannopoulos12}\cite{aviram16} (some papers have more than one attack). These attacks represent all of the notable published downgrade attacks that we are aware of, starting from the first version of TLS (SSL~2.0) until draft-10 of the upcoming version (TLS~1.3). We summarise them in section \ref{downgrade-attacks}. Second, we extract the features that characterise each attack (which we refer to as vectors), namely: the attacker targets an element that defines the mode of the protocol which can be the protocol algorithms, version, or the TLS layer, in order to modify or remove. The attacker also needs to exploit a vulnerability, which can be due to implementation, design, or trust-model. The downgrade is achieved by using a method which can be message modification, dropping, or injection. Finally, the attack results in a damage which can be either broken security or weakened security. These four main vectors are intrinsic to any downgrade attack under the specified attacker model and can therefore be used to characterise each attack in that model. Third, after identifying the vectors, we devise the taxonomy. We define the notions of the taxonomy's categories and sub-categories in section \ref{taxonomy}. Finally, we show the taxonomy's application in classifying known TLS downgrade attacks. 

\section{Downgrade Attacks in TLS, a Brief Survey} \label{downgrade-attacks}
In this section, we briefly survey the TLS downgrade attacks that we have analysed in order to devise the taxonomy. We highlight the attack names in \textbf{Bold} and we use these names throughout the paper. We assume the reader's familiarity with the TLS technical details. The unfamiliar reader is advised to read Appendix \ref{appendix:a}, which provides the required background to comprehend the rest of the paper. 

Downgrade attacks have existed since the very early versions of TLS: SSL~2.0 \cite{sslv2} and SSL~3.0 \cite{sslv3}. SSL~2.0 suffers from the \textbf{\quotes{ciphersuite rollback}} attack, where the attacker limits SSL~2.0 strength to the \quotes{least common denominator}, i.e. the weakest ciphersuite, by modifying the ciphersuites list in one or both of the \texttt{Hello} messages that both parties exchange so that they offer the weakest ciphersuite \cite{wagner96}\cite{turner11}, e.g. export-grade or \quotes{NULL} encryption ciphersuites. To mitigate such attacks, SSL~3.0 mandated a MAC of the protocol's transcript in the \texttt{Finished} messages which needs to be verified by both parties to ensure identical views of the transcript (i.e. unmodified messages). \par 

However, SSL~3.0 is vulnerable to the \textbf{\quotes{version rollback}} attack that works by modifying the client's proposed version from SSL~3.0 to SSL~2.0 \cite{wagner96}. This in turn leads SSL~3.0 servers that support SSL~2.0 to fall back to SSL~2.0. Hence, all SSL~2.0 weaknesses will be inherited in that handshake including the lack of integrity and authentication checks for the protocol's transcript as we described above, which render the downgrade undetected. \par 

Another design flaw in SSL~3.0 allows a theoretical attack named the \textbf{\quotes{key-exchange rollback}} attack, which is a result of lack of authentication for the server's selected ciphersuite (which includes the name of the key-exchange algorithm) before the \texttt{Finished} MACs \cite{wagner96}. In this attack, the attacker modifies the client's proposed key-exchange algorithm from RSA to DHE, which makes the communicating parties have different views about the key-exchange algorithm. That is, the server sends DHE key parameters in the \texttt{ServerKeyExchange} message while the client treats them according to export-grade RSA algorithm. These mismatched views about the key-exchange result in generating breakable keys which are then used by the attacker to forge the \texttt{Finished} MACs to hide the attack, impersonate each party to the other, and to decrypt the application data. \par

In \cite{Mavrogiannopoulos12}, an attack which we call the \textbf{\quotes{DHE key-exchange rollback}} is presented. It can be considered a variant of the \textbf{\quotes{key-exchange rollback}} in \cite{wagner96}. In this attack the attacker modifies the client's proposed key-exchange algorithm from DHE to ECDHE. As a result, the server sends a \texttt{ServerKeyExchange} that contains ECDHE parameters based on the client offer while the client treats them as DHE parameters. The client does not know the selected key-exchange algorithm by the server since the selected ciphersuite (which includes the key-exchange algorithm) is not authenticated in the \texttt{ServerKeyExchange}. Similar to the \textbf{\quotes{key-exchange rollback}} attack in \cite{wagner96}, these mismatched views about the key-exchange algorithm result in breakable keys, which allow the attacker to recover the pre-master and master secretes. Consequently, be able to forge the \texttt{Finished} MACs to hide the modifications in the \texttt{Hello} messages, impersonate each party to the other, and decrypt the application data.  

Version downgrade is not exclusive to SSL~3.0. The Padding Oracle On Downgraded Legacy Encryption \textbf{(POODLE)} attack \cite{moller14} shows the possibility of version downgrade in recent versions of TLS (up to TLS~1.2) by exploiting the \quotes{downgrade dance}, a client-side implementation technique that is used by some TLS clients (e.g. web browsers). It makes the client fall back to a lower version and retries the handshake if the initial handshake failed for any reason \cite{moller14}. In the POODLE attack, a man-in-the-middle abuses this mechanism by dropping the \texttt{ClientHello} to lead the client to fall back to SSL~3.0. This in turn brings the specific flaw that is in the CBC padding in all block ciphers in SSL~3.0, which allows the attacker to decrypt some of the SSL session's data such as the cookies that may contain login passwords.\par

In \cite{adrian15}, the \textbf{Logjam} attack is presented. It uses a method similar to the one we explained in the illustrative example in section \ref{example}. The Logjam attack is applicable to DHE key-exchange. It works by modifying the \texttt{Hello} messages to misguide the server into selecting an export-grade DHE ciphersuite which result in weak DHE keys. As stated earlier, TLS up to version~1.2 does not authenticate the server's selected ciphersuite (which includes the key-exchange algorithm) until the \texttt{Finished} MACs. As a result, the client receives weak key parameters and generates weak keys based on the server's weak parameters. The lack of early authentication of the server's selected ciphersuite gives the attacker a window of time to recover the master secret from the weakly generated keys in real-time, before the \texttt{Finished} MACs. Consequently, the attacker can forge the \texttt{Finished} MACs to hide the modifications in the \texttt{Hello} messages, and decrypt the the application data. \par

A similar attack called the Factoring RSA Export Keys \textbf{(FREAK)} attack \cite{beurdouche15} is performed using a method similar to the one used in the Logjam attack \cite{adrian15}, which leads the server into selecting an export-grade ciphersuite. However, FREAK is applicable to RSA key-exchange and requires a client implementation vulnerability that makes a client that does not support export-grade ciphersuites accept a \texttt{ServerKeyExchange} message with weak ephemeral export-grade RSA key parameters, while the key-exchange algorithm is RSA (note that the \texttt{ServerKeyExchange} message must not be sent when the key-exchange algorithm is non-export-grade RSA \cite{tls12}). However, the \texttt{ServerKeyExchange} is sent in export-grade RSA or in (EC)DHE key-exchange. This implementation vulnerability leads the client to use the export-grade RSA key parameters that are provided in the \texttt{ServerKeyExchange} to encrypt the pre-master secret instead of encrypting it with the long-term (presumably strong) RSA key that is provided in the server's \texttt{Certificate}. This results in breakable keys that can be used to forge the \texttt{Finished} MACs and decrypt the application data. \par

In \cite{aviram16}, a variant of the Decrypting RSA using Obsolete and Weakened eNcryption (\textbf{DROWN}) attack (the \quotes{special DROWN}) that exploits an OpenSSL server implementation bug \cite{cve15} is presented. The attack enables a man-in-the-middle to force a client and server into choosing RSA key-exchange algorithm despite their preference for non-RSA (e.g. (EC)DHE) by modifying the \texttt{Hello} messages. The attacker then make use of a known flaw that can be exploited if the server's RSA key is shared with an SSLv2 server using an attack called Bleichenbacher attack \cite{ble98} which enables the attacker to recover the plaintext of an RSA encryption (i.e. the pre-master secret) by using the SSLv2 server as a decryption oracle. If the attacker can break the pre-master secret, he can break the master secret and forge the \texttt{Finished} MACs to hide the attack, and be able to decrypt the application data. 

Another case of downgrade attack is the \textbf{\quotes{Forward Secrecy rollback}} attack \cite{beurdouche15}, in which the attacker exploits an implementation vulnerability to make the client fall back from Forward Secrecy (FS)\footnote{Forward Secrecy (FS) is a property that guarantees that a compromised long-term key does not compromise past session keys \cite{menezes96})} mode to non-FS mode by dropping the \texttt{ServerKeyExchange} message. However, non-FS mode does not result in immediate breakage of any security guarantee such as secrecy unless the long-term key that encrypts the session keys got broken after the session keys have been used to encrypt application data. \par

In \cite{bhargavan2016transcript}, a downgrade attack in TLS~1.0 and TLS~1.1 is illustrated. The attack comes under a family of attacks named Security Losses from Obsolete and Truncated Transcript Hashes (\textbf{SLOTH}). This attack is possible due to the use of non collision resistant hash functions (MD5 and SHA-1) in the \texttt{Finished} MACs. The use of MD5 and SHA-1 is mandated by the TLS~1.0-1.1 specifications \cite{tls1}\cite{tls11}. Non collision resistant hash functions allow the attacker to modify the \texttt{Hello} messages without being detected in the \texttt{Finished} MACs by creating a prefix-collision in the transcript hashes \cite{bhargavan2016transcript}. \par

Downgrade attacks in multi-layered protocols that negotiate upgrading the connection to operate over TLS have been shown to be prevalent based on an empirical analysis of SMTP deployment in the IPv4 internet space \cite{durumeric15}. In \cite{durumeric15} they found evidence for corrupted \texttt{STARTTLS} commands which downgrade \textbf{SMTPS to SMTP} in more than 41,000 mail servers. \par

Similarly, downgraded TLS as a result of \textbf{proxied HTTPS} connections\footnote{A proxy refers to an entity that is located between the client and server and splits the TLS session into two separate sessions. As a result, the client encrypts the data using the proxy's public-key.} has been shown to be prevalent. In \cite{durumeric17}, empirical data show that 10-40\% of the proxied TLS connections advertise known broken cryptographic choices \cite{durumeric17}. \par

Downgrade attacks continued to appear until draft-10 of the coming version of TLS (TLS~1.3 \cite{tls13rev10}), where \cite{bhargavan2016downgrade} report three possible downgrade attacks in TLS~1.3 draft-10. The first attack is similar in spirit to SSL~3.0 \textbf{\quotes{version rollback}} attack that we explained earlier in this section. In this attack, the attacker modifies the proposed version to TLS~1.2 and enjoys the vulnerabilities in TLS~1.2 that (in the presence of export-grade ciphersuites either on the server side or in both sides) enable him to break the master secret before the \texttt{Finished} MACs as in \cite{adrian15}\cite{beurdouche15}, hence circumventing downgrade detection. \par

The second attack in TLS~1.3 draft-10, which we call the \quotes{\textbf{downgrade dance version rollback}} attack, employs a method similar to the one employed in the POODLE attack \cite{moller14}, i.e. the attacker drops the initial handshake message one or more times to lead the clients that implement the \quotes{downgrade dance} mechanism to fall back to a lower version such as TLS~1.2, hence circumvent detection due to downgrade security weaknesses in TLS~1.2 and lower versions. \par

Finally, the third reported downgrade attack in TLS~1.3 draft-10, which we call the \quotes{\textbf{\texttt{HelloRetry} downgrade}} attack, occurs when an attacker injects a \texttt{HelloRetryRequest} message to downgrade the (EC)DHE group to a less preferred group despite the client and server preference to use another group. This attack can circumvent detection because the transcript hash restarts with every \texttt{HelloRetryRequest} \cite{bhargavan2016downgrade}. However, consequent TLS~1.3 drafts mitigated this attack by continuing the hashes over retries \cite{bhargavan2016downgrade}.  
    
\section{Taxonomy of Downgrade Attacks} \label{taxonomy}
\begin{figure*}[!tp]
\centering
\includegraphics[width=\textwidth]{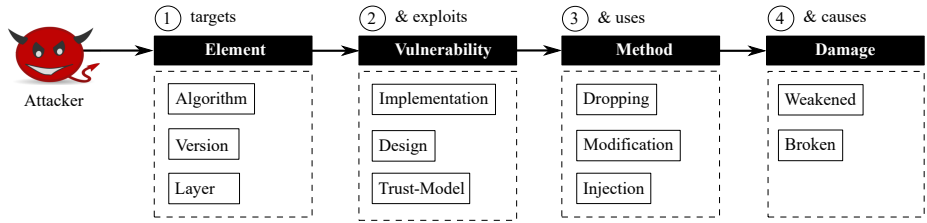}
\caption{A taxonomy of downgrade attacks in the TLS protocol and application protocols using TLS.}
\label{fig:taxonomy}
\end{figure*}

Based on the surveyed attacks in section \ref{downgrade-attacks}, we distill four vectors that characterise the surveyed downgrade attacks, namely: element, vulnerability, method, and damage. These vectors represent the taxonomy's main categories. We define the notions of the categories and sub-categories that we use in our taxonomy. Figure \ref{fig:taxonomy} summarises the taxonomy. 

\begin{enumerate}
\item \textbf{Element:} The element refers to the protocol element that is being negotiated between the communicating parties. The element's value is intrinsic in defining the protocol mode, i.e. the security level of the protocol run. The element is targeted by the attacker because either modifying or removing it will result in either a less secure, non secure, or less preferred mode of the protocol. We categorise the element into three sub-categories as follows: 
	\begin{enumerate}
	\item Algorithm: The algorithm refers to the cryptographic algorithms, e.g. key-exchange, encryption, hash, signature, etc. and their parameters such as block cipher modes of operation and key lengths, that are being negotiated to be used in subsequent messages of the protocol. Generally, in TLS, the main algorithms are represented by the ciphersuite, but they can also be represented by other parameters that are not part of the ciphersuite such as the extensions.
			
	\item Version: The version refers to the protocol version. A number of protocols including TLS allow their communicating parties to support multiple versions, negotiate the protocol version that both communicating parties will run, and allow them to fall back to a lower version to match the other party's version if the versions at both ends do not match.

	\item Layer: The layer refers to the whole TLS layer which is negotiated and optionally added in some legacy protocols. In such protocols like SMTP \cite{smtp2001} for example, TLS encapsulation is negotiated through specific upgrade messages, e.g. STARTTLS \cite{starttls2002}, in order to upgrade the protocol from an insecure (plaintext and unauthenticated) to a secure (encrypted and/or authenticated) mode.
	\end{enumerate}

\item \textbf{Vulnerability:} Like any attack performed by an external man-in-the-middle, downgrade attacks require a vulnerability to be exploited. We categorise the vulnerability into three sub-categories as follows:
	\begin{enumerate} 
	\item Implementation: An implementation vulnerability refers to a faulty protocol implementation. The existence of implementation vulnerabilities can be due to various reasons, for example, a programmer's fault, a state-machine bug, or a malware that corrupted the code. 
	  
	\item Design: A design vulnerability refers to a flaw in the protocol design (i.e. the specifications). The protocol design is independent of the implementation. That is, even if the protocol was perfectly implemented, an attacker can exploit a design flaw to perform a downgrade attack. 

	\item Trust-Model: A trust-model vulnerability refers to a flaw in the architectural aspect (the TLS ecosystem in our case) and the trusted parties involved in this architecture which is independent of the protocol design and implementation.  
	\end{enumerate}
	
\item \textbf{Method:} The method refers to the method used by the attacker to perform the downgrade. We categorise the method into three sub-categories as follows: 
	\begin{enumerate}
	\item Modification: In the modification method, the attacker modifies the content of one or more protocol messages that negotiate the element (i.e. algorithm, version, layer). If the protocol does not employ any integrity nor authentication checks for the handshake transcript, the downgrade attack can be trivially performed. Otherwise, the attacker needs to find ways to circumvent the checks, for example, break the master secret or create colliding hashes for the transcript.  

	\item Dropping: In the dropping method, the attacker drops one or more protocol messages (possibly more than once). 

	\item Injection: In the injection method, the attacker sends a new message to one of the communicating parties by impersonating the party's peer, for example to request a different algorithm or version than what is initially offered by the communicating party. The injection method is trivial in the absence of transcript integrity and authentication checks. Otherwise, it requires circumventing the integrity and authentication checks. 
	\end{enumerate}

\item \textbf{Damage:} The damage refers to the resulted damage after a successful downgrade attack. We categorise the damage into two sub-categories as follows: 
	\begin{enumerate}
	\item Broken Security: Broken security refers to downgrade attacks that result in allowing the attacker to break one or more main security goals that the protocol claims to guarantee. In TLS the guarantees are: secrecy, authentication, and integrity. 

	\item Weakened Security: Unlike the broken security damage, weakened security does not result in immediate breakage of any of the main security guarantees. Instead, weakened security refers to attacks that result in making the communicating parties choose a non-recommended or less preferred mode, which is not broken yet. 
	\end{enumerate}

\end{enumerate}

\section{Discussion} \label{discussion}
In Table \ref{table:taxonomy-application}, we show the taxonomy's application in classifying the surveyed TLS downgrade attacks. Then we discuss our reasoning in some of the noteworthy cases (we will refer to the attacks by their reference number according to the numbers in Table \ref{table:taxonomy-application}). 

\begin{table*}[!t] 
\centering 
\begin{threeparttable}
\resizebox{0.90\textwidth}{!} {
\begin{tabular} {ll*{18}l}
\toprule
\multicolumn{1}{c}{No.} 
&& \multicolumn{1}{c}{Attack} 
&&\multicolumn{3}{c}{Element}    
&&\multicolumn{3}{c}{Vuln.}  
&&\multicolumn{3}{c}{Method}  
&&\multicolumn{2}{c}{Damage}  
\\
\cmidrule(lr){1-1} 
\cmidrule(lr){3-3}
\cmidrule(lr){5-7} 
\cmidrule(lr){9-11}
\cmidrule(lr){13-15}
\cmidrule(lr){17-18}
& \vline 
& & \vline &\rot{Algorithm} & \rot{Version} & \rot{Layer} 
& \vline & \rot{Implementation} & \rot{Design} & \rot{Trust-model} 
& \vline & \rot{Dropping} & \rot{Modification} &\rot{Injection} 
& \vline & \rot{Weakened} & \rot{Broken} 
\\ 
\midrule
\rowcolor[gray]{.9} 
01 & \vline  & SSL~2.0 Ciphersuite rollback \cite{wagner96}\textsuperscript{\textbf{*}} &\vline  & \cmark & {}     & {}     & \vline  & {}     & \cmark  & {}      & \vline  & {} &\cmark  & & \vline &    & \cmark    \\

02 & \vline & SSL~3.0 Version rollback \cite{wagner96}\textsuperscript{\textbf{*}}     &\vline  & {}     & \cmark & {}     & \vline  & {}     & \cmark   & {}      & \vline & {} & \cmark & & \vline & {} & \cmark   \\

\rowcolor[gray]{.9}
03 & \vline & SSL~3.0 key-exchange rollback \cite{wagner96}\textsuperscript{\textbf{*}}& \vline & \cmark & {}     & {}     & \vline  & {}     & \cmark   & {}      & \vline & {} &\cmark  & &\vline & {} & \cmark    \\

04 & \vline & DHE key-exchange rollback \cite{Mavrogiannopoulos12} & \vline & \cmark & {}     & {}     & \vline  & {}     & \cmark   & {}      & \vline & {} &\cmark  & &\vline & {} & \cmark    \\

\rowcolor[gray]{.9}
05 & \vline & TLS~1.0-1.1 SLOTH \cite{bhargavan2016transcript}\textsuperscript{\textbf{*}} & \vline & & \cmark    & {}     & \vline   & {}     & \cmark   & {}      & \vline & {} &\cmark  & &\vline & {} & \cmark  \\
 
06 & \vline & POODLE version downgrade \cite{moller14}      & \vline & {}     & \cmark & {}    & \vline   & \cmark  & {}      & {}      & \vline & \cmark & {} & &\vline & {} & \cmark     \\

\rowcolor[gray]{.9} 
07 & \vline & FREAK \cite{beurdouche15}                           & \vline & \cmark & {}    & {}     & \vline   & \cmark  & {}      & {}      & \vline & {}     & \cmark & & \vline & {} & \cmark   \\

08 & \vline & DROWN \cite{aviram16}  & \vline & \cmark & {}    & {}     & \vline   & {}  & {}      & \cmark      & \vline & {}     & \cmark & & \vline & {} & \cmark   \\

\rowcolor[gray]{.9}
09 & \vline & Forward Secrecy rollback \cite{beurdouche15}        & \vline & \cmark & {}    & {}     & \vline   & \cmark  & {}      & {}       & \vline & \cmark & {}    & & \vline & \cmark  & {}  \\

10 & \vline & Logjam \cite{adrian15}                         & \vline & \cmark & {}   & {}     & \vline    & {}     & \cmark   & {}      & \vline  & {}     &\cmark & & \vline & {} & \cmark   \\

\rowcolor[gray]{.9}
11 & \vline & SMTPS to SMTP \cite{durumeric15}\textsuperscript{\textbf{*}} & \vline & {}     & {}     & \cmark & \vline &   & \cmark  & {}       & \vline  & {}     &\cmark & & \vline & {} & \cmark \\

12 & \vline & Proxied HTTPS \cite{durumeric17}\textsuperscript{\textbf{*}}       & \vline &      &      & \cmark & \vline & {}      & {}      & \cmark   & \vline  & {}     &{} & \cmark &\vline & {} & \cmark  \\

\rowcolor[gray]{.9}
13 & \vline & TLS~1.3 Version rollback \cite{bhargavan2016downgrade}\textsuperscript{\textbf{*}} &\vline  & {}     & \cmark & {}     & \vline  & {}     & \cmark   & {}      & \vline & {} & \cmark & & \vline & {} & \cmark \\

14 & \vline & TLS~1.3 Downgrade-dance version fallback \cite{bhargavan2016downgrade}\textsuperscript{\textbf{*}} & \vline & {}     & \cmark & {}    & \vline   & \cmark  & {}      & {}      & \vline & \cmark & {} & &\vline & {} & \cmark   \\

\rowcolor[gray]{.9}
15 & \vline & TLS~1.3 \texttt{HelloRetry} downgrade \cite{bhargavan2016downgrade}\textsuperscript{\textbf{*}} & \vline & \cmark  & {} & {}    & \vline   & {}  & \cmark & {}  & \vline & {} & {} &\cmark &\vline  & \cmark & {} \\
\bottomrule
\\

\end{tabular}
}
\end{threeparttable}
\caption{Classifying the surveyed downgrade attacks using our taxonomy. Attacks that are followed by \quotes{\textbf{*}} do not have an implementation and are either theoretical or based on evidence from measurement studies.}
\label{table:taxonomy-application}
\end{table*}
It should be noted that classifying attacks that have implementation is straightforward as is the case in the attacks: \textbf{04}, \textbf{06}, \textbf{07}, \textbf{08}, \textbf{09}, and \textbf{10} where their classifications in Table \ref{table:taxonomy-application} are self-explanatory based on mapping the surveyed attacks description in section \ref{downgrade-attacks} with the categories in Table \ref{table:taxonomy-application}. On the other hand, classifying either theoretical attacks such as \textbf{01}, \textbf{02}, \textbf{03}, \textbf{05}, \textbf{13}, \textbf{14}, and \textbf{15}, or attacks that have been reported based on evidence from empirical data such as \textbf{11} and \textbf{12}, is less straightforward and requires making some assumptions. \par 

Ideally the taxonomy helps in classifying concrete attacks that have implementation. However, for the sake of illustration, we make some assumptions (mostly worst case assumptions) to mimic a concrete attack case from the general attack that does not have an implementation. In the following, we elaborate more on these cases. 
  
Attacks \textbf{01}, \textbf{02}, and \textbf{03} are theoretical. We classify the damage on these attacks based on the worst case assumption as follows: In \textbf{01}, we assume that the attacker can select export-grade or \quotes{NULL} encryption ciphersuites, which breaks a main security guarantees of TLS. In \textbf{02}, once the attacker downgrades SSL~3.0 to SSL~2.0, he can perform attack \textbf{01} without being detected due to lack of downgrade security in SSL~2.0. In \textbf{03}, we assume that the attacker can break the master secret. Similar to the FREAK \cite{beurdouche15} and Logjam \cite{adrian15} attacks, this allows the attacker to forge the \texttt{Finished} MACs which enables him to impersonate the client and/or the server and decrypt the application data, and this breaks main security guarantees. \par   
 
Attack \textbf{05} is a theoretical attack. Based on the worst case assumption, we classify the downgraded element under the version element. The attacker can modify the version as well as the algorithms and hide the attack by producing prefix collision in the transcript hashes which are computed using non collision resistant hashes (MD-5 and SHA1 based on the protocol design and specifications) that go into the \texttt{Finished} MACs. If the attacker succeeded in downgrading the version to a broken version such as SSL~3.0, he can break main security guarantees (e.g. the CBC flaw in the symmetric encryption in SSL~3.0), hence the damage in attack \textbf{05} is classified under broken category.\par

Although attack \textbf{08} has an implementation but it is quite complex attack and its vulnerability classification is noteworthy. We classified its vulnerability under the trust model. By contemplating the main cause that allows this downgrade attack to succeed we find the main reason lies in breaking the the pre-master the master secret that is then used to forge the \texttt{Finished} MACs, otherwise the attack will be detected. In this attack, the attacker can decrypt the pre-master secret if it is encrypted with an RSA key (even a strong 2048-bit RSA key), if the key is shared with an SSLv2 server (e.g. both servers uses the same certificate). Sharing RSA keys among servers is a trust-model vulnerability that allows the key sharing, rather than a protocol design nor implementation.
 
Attack \textbf{11} is based on evidence from real-world deployment. Based on the reported evidence described in \cite{durumeric15}, the method is classified under modification. However, dropping can also work as another method based on the STARTTLS specifications \cite{starttls2002}. Since forcing TLS is not mandated by the SMTP protocol design and specifications, we do not consider the \quotes{fail open} local policy as an implementation vulnerability but a design one. \par 

Attack \textbf{12} is widely known as HTTPS interception, where a man-in-the-middle (represented by a proxy) has full control over the TLS channel, which gives him the ability to downgrade TLS (algorithm, version, or layer). The empirical results in \cite{durumeric17} shows an evidence of downgraded TLS version and algorithm due to proxied HTTPS. However, in fact, the man-in-the-middle can send the client's data to the server in cleartext. Therefore, based on the worst case assumption, the targeted element is classified under layer. The method is classified under injection since the man-in-the-middle injects a new message to the server by impersonating the client. \par

Attack \textbf{13} is similar in spirit to \textbf{02} that occurs in SSL~3.0 which is due to a design vulnerability. In TLS~1.3, the attack has been mitigated by redesigning the server's nonce to signal the received client's version \cite{tls13rev25}. \par

Attack \textbf{14} is similar in spirit to \textbf{06} which targets the protocol version. If the attacker succeed in downgrading the version to a flawed version that has downgrade security weaknesses (as is the case in TLS~1.2 and below), the attacker can break main security guarantees based on the worst case assumption. \par

Attack \textbf{15} damage is classified under weakened security because as of this writing, no known broken (EC)DHE group elements are allowed in TLS 1.3 by design. Therefore, under the worst case assumption, the resulted damage leads both parties to agree on the least preferred DHE group. \par

Finally, as Table \ref{table:taxonomy-application} shows, in most of the cases the resulted damage is broken security except in two cases. 

\section{Conclusion and Future Work} \label{conclusion}
In conclusion, we introduce the first taxonomy of downgrade attacks in the TLS protocol and application protocols using TLS. Our taxonomy classifies downgrade attacks with respect to four vectors: element, vulnerability, method, and damage. It is based on a through analysis of fifteen TLS downgrade attack cases under the assumption of an external man-in-the-middle attacker model. In addition, we provided a brief survey of all notable published TLS downgrade attacks to date. Finally, we demonstrate our taxonomy's application in classifying known TLS downgrade attacks. For future work, we plan to test the taxonomy on downgrade attacks in protocols other than TLS for potential generalisation of the taxonomy. Furthermore, we believe that the taxonomy has the potential of serving as a useful tool in devising downgrade attack severity assessment model, which can enable ranking the attack severity, which can help in identifying the attacks that require more research efforts to mitigate them.   

\section{Acknowledgment} \label{acknowledgment}
The authors would like to thank Prof. Kenny Paterson, Prof. Andrew Martin, and Nicholas Moore for their feedback, and Mary Bispham, Ilias Giechaskiel, Jacqueline Eggenschwiler, and John Gallacher for proofreading earlier versions of this paper.

\FloatBarrier
\bibliographystyle{splncs03}
\bibliography{refs}
\begin{subappendices}
\renewcommand{\thesection}{\Alph{section}}
\section{The TLS Protocol}\label{appendix:a}
\subsection{TLS, a General Overview}
The main goal of TLS is to provide a secure communication channel between two communicating parties \cite{tls12}, ideally client (initiator $I$) and server (responder $R$). TLS consists of two sub-protocols: the handshake protocol and the record protocol \cite{tls12}. Briefly, the handshake protocol is responsible for version and ciphersuite negotiation, client and server authentication, and key exchange. On the other hand, the record protocol is responsible for carrying the protected application data, encrypted with the just negotiated keys in the handshake. As of this writing, TLS~1.2 \cite{tls12} is the currently deployed standard. The coming version of TLS, TLS~1.3 \cite{tls13rev25}, is still work in progress. Figure \ref{fig:TLS-ECDHE} shows the message sequence diagram for TLS~1.2 using Ephemeral Diffie-Hellman (EC)DHE\footnote{We use (EC)DHE as an abbreviation for: Elliptic-Curve Ephemeral Diffie-Hellman (ECDHE) or Ephemeral Diffie-Hellman (DHE).} key-exchange \cite{diffie76}, Figure \ref{fig:TLS-RSA} shows TLS~1.2 using Rivest-Shamir-Adleman (RSA) key-exchange \cite{rivest78}, and Figure \ref{fig:TLS13} illustrates the changes in the \texttt{Hello} messages in TLS~1.3 based on the latest draft (draft-25 as of this writing) \cite{tls13rev25}. Our scope in this paper is TLS in certificate-based unilateral server-authentication mode. In the diagrams, the messages are represented by their initials (e.g. \texttt{CH} refers to \texttt{ClientHello}). Throughout the paper, the protocol messages are distinguished by a \texttt{TypeWriter} font. 

\subsection{TLS~1.2 Handshake Protocol}
\begin{figure}[!tp]
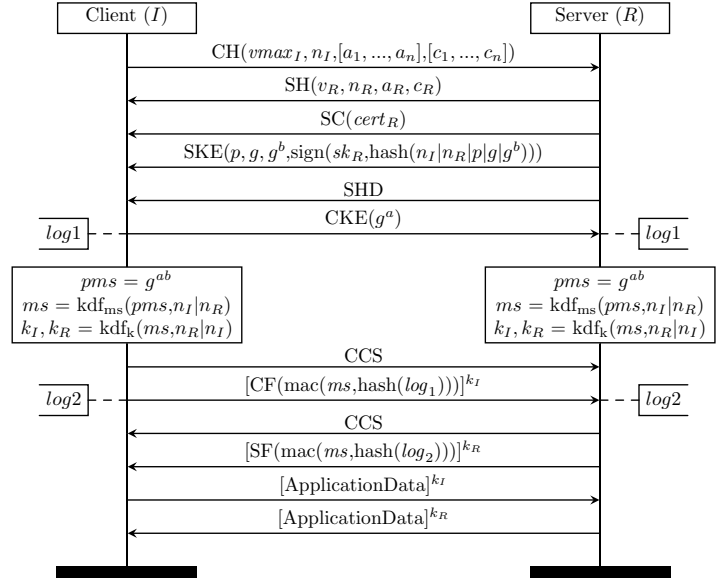
 
\centering
\resizebox{0.8\textwidth}{!}{
\setmsckeyword{} 
\drawframe{no} 

\begin{msc}[normal values, /msc/level height=0.6cm, /msc/label distance=0.5ex, /msc/first level height=0.6cm, /msc/last level height=0.6cm]{} 
\setlength{\instwidth}{2.5\mscunit} 
\setlength{\instdist}{6\mscunit} 
\declinst{I}{}{Client ($I$)}%
\declinst{R}{}{Server ($R$)}%

\mess{CH($\mathit{vmax}_I,n_I,$[$a_1,...,a_n$],[$c_1,...,c_n$])} {I}{R}
\nextlevel

\mess{SH($v_R,n_R,a_R,c_R$)} {R}{I}
\nextlevel

\mess{SC($\mathit{cert}_R$)} {R}{I}

\nextlevel
\mess{SKE($p,g,g^b$,sign($\mathit{sk}_R$,hash($n_I|n_R|p|g|g^b$)))} {R}{I}

\nextlevel
\mess{SHD}{R}{I}

\nextlevel
\mess{CKE($g^a$)}{I}{R}
\msccomment[msccomment distance=0.7cm]{$log1$}{I}
\msccomment[msccomment distance=0.7cm,side=right]{$log1$}{R}

\nextlevel

\action*{\parbox{4cm} {\centering $pms$ = $g^{\mathit{ab}}$ \\ $ms$ = kdf\textsubscript{ms}($pms$,$n_I|n_R$) \\ $k_{I},k_{R}$ = kdf\textsubscript{k}($ms$,$n_R|n_I$)}}{I}
\action*{\parbox{4cm} {\centering $pms$ = $g^{\mathit{ab}}$ \\ $ms$ = kdf\textsubscript{ms}($pms$,$n_I|n_R$) \\ $k_{I},k_{R}$ = kdf\textsubscript{k}($ms$,$n_R|n_I$)}}{R}
\nextlevel[3]

\mess{CCS} {I}{R}
\nextlevel
\mess{[CF(mac($\mathit{ms}$,hash($\mathit{log}_1$)))]$^{k_I}$} {I}{R}
\msccomment[msccomment distance=0.7cm]{$log2$}{I}
\msccomment[msccomment distance=0.7cm,side=right]{$log2$}{R}

\nextlevel
\mess{CCS} {R}{I}
\nextlevel
\mess{[SF(mac($\mathit{ms}$,hash($\mathit{log}_2$)))]$^{k_R}$} {R}{I}

\nextlevel
\mess{[ApplicationData]$^{k_I}$} {I}{R}

\nextlevel
\mess{[ApplicationData]$^{k_R}$} {R}{I}

\end{msc}
} 

\caption{Message sequence diagram for TLS~1.2 with (EC)DHE key-exchange.}
\label{fig:TLS-ECDHE} 
\end{figure}

We briefly describe the TLS~1.2 handshake protocol in certificate-based unilateral server-authentication mode based on the Internet Engineering Task Force (IETF) standard's specifications \cite{tls12}. A detailed description of the protocol can be found in \cite{tls12}. As depicted in Figure \ref{fig:TLS-ECDHE}, the handshake protocol works as follows: First, the client sends a \texttt{ClientHello} (\texttt{CH}) message to initiate a connection with the server. This message contains: the maximum version of TLS that the client supports ($\mathit{vmax}_I$); the client's random value ($n_I$); optionally, a session identifier if the session is resumed ($\mathit{session}_{\mathit{ID}}$); a list of ciphersuites that the client supports ordered by preference ($[a_1,...,a_n]$); a list of compression methods that the client supports ordered by preference ($[c_1,...,c_n]$); and finally, an optional list of extensions ($[e_1,...,e_n]$). \par

Second, the server responds with a \texttt{ServerHello} (\texttt{SH}) message. This message contains: the server's selected TLS version ($v_R$); the server's nonce ($n_R$); optionally, a session identifier in case of session resumption ($\mathit{session}_{\mathit{ID}}$); the selected ciphersuite based on the client's proposed list ($a_R$); the selected compression method from the client's proposed list ($c_R$); and optionally, a list of the extensions that are requested by the client and supported by the server ($[e_1,...,e_n]$). After that, the server sends a \texttt{ServerCertificate} (\texttt{SC}), which contains the server's certificate ($\mathit{cert}_R$) if server authentication is required. Then, if the key-exchange algorithm is (EC)DHE (see \cite{diffie76} for details about the DH algorithm), the server sends a \texttt{ServerKeyExchange} (\texttt{SKE}) message. This message must not be sent when the key-exchange algorithm is RSA (see \cite{rivest78} for details about the RSA algorithm). The \texttt{ServerKeyExchange} contains the server's (EC)DHE public-key parameters and a signature over a hash of the nonces ($n_I$ and $n_R$) and the (EC)DHE key parameters. In case of DHE (i.e. Finite Field DHE), the key parameters are: the prime ($p$), the generator ($g$), and the server's public value ($g^b$). We omit describing the ECDHE parameters and we refer the reader to \cite{ecc06} for details about ECDHE key parameters. Finally, the server sends a \texttt{ServerHelloDone} (\texttt{SHD}) to indicate to the client that it finished its part of the key-exchange. \par

\begin{figure}[!tp]
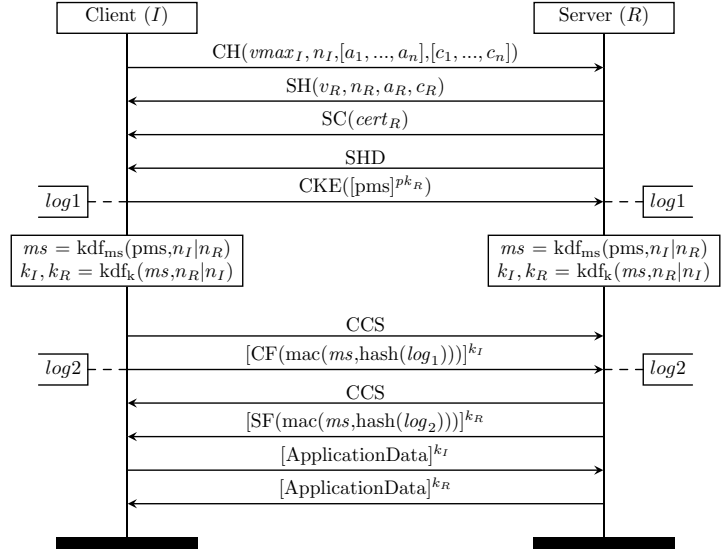
 %
\centering
\resizebox{0.8\textwidth}{!}{%
\setmsckeyword{} 
\drawframe{no} 
\begin{msc}[normal values, /msc/level height=0.6cm, /msc/label distance=0.5ex, /msc/environment distance=0cm, /msc/first level height=0.6cm, /msc/last level height=0.6cm]{} 
\setlength{\instwidth}{2.5\mscunit} 
\setlength{\instdist}{6\mscunit} 

\declinst{I}{}{Client ($I$)}%
\declinst{R}{}{Server ($R$)}%

\mess{CH($\mathit{vmax}_I,n_I,$[$a_1,...,a_n$],[$c_1,...,c_n$])} {I}{R}
\nextlevel

\mess{SH($v_R,n_R,a_R,c_R$)} {R}{I}
\nextlevel

\mess{SC($\mathit{cert}_R$)} {R}{I}

\nextlevel
\mess{SHD}{R}{I}

\nextlevel
\mess{CKE($[\mathrm{pms}]^{\mathit{pk}_R}$)}{I}{R}
\msccomment[msccomment distance=0.7cm]{$log1$}{I}
\msccomment[msccomment distance=0.7cm,side=right]{$log1$}{R}

\nextlevel
\action*{\parbox{4cm} {\centering $\mathit{ms}$ = kdf\textsubscript{ms}($\mathrm{pms}$,$n_I|n_R$) \\ $k_{I},k_{R}$ = kdf\textsubscript{k}($\mathit{ms}$,$n_R|n_I$)}}{I}
\action*{\parbox{4cm} {\centering $\mathit{ms}$ = kdf\textsubscript{ms}($\mathrm{pms}$,$n_I|n_R$) \\ $k_{I},k_{R}$ = kdf\textsubscript{k}($\mathit{ms}$,$n_R|n_I$)}}{R}
\nextlevel[3]

\mess{CCS} {I}{R}
\nextlevel
\mess{[CF(mac($\mathit{ms}$,hash($\mathit{log}_1$)))]$^{k_I}$} {I}{R}
\msccomment[msccomment distance=0.7cm]{$log2$}{I}
\msccomment[msccomment distance=0.7cm,side=right]{$log2$}{R}

\nextlevel
\mess{CCS} {R}{I}
\nextlevel
\mess{[SF(mac($\mathit{ms}$,hash($\mathit{log}_2$)))]$^{k_R}$} {R}{I}

\nextlevel
\mess{[ApplicationData]$^{k_I}$} {I}{R}

\nextlevel
\mess{[ApplicationData]$^{k_R}$} {R}{I}
\end{msc}
} 
\caption{Message sequence diagram for TLS~1.2 with RSA key-exchange.}
\label{fig:TLS-RSA}
\end{figure}

Third, upon receiving the \texttt{ServerHelloDone} the client should verify the server's certificate and the compatibility of the server's selected parameters in the \texttt{ServerHello}. After that, the client sends a \texttt{ClientKeyExchange} (\texttt{CKE}) to set the pre-master secret. The content of the \texttt{ClientKeyExchange} depends on the key-exchange algorithm. If the key-exchange algorithm is RSA, the  client sends the pre-master secret encrypted with the server's long-term RSA public-key ($[\mathit{pms}]^{\mathit{pk}_R}$) as illustrated in Figure  \ref{fig:TLS-RSA}. If the key-exchange algorithm is DHE, the client sends its DHE public value ($g^a$) to allow the server to compute the shared DHE  secret-key ($g^{\mathit{ab}}$) as illustrated in Figure \ref{fig:TLS-ECDHE}. After that, both parties compute the master secret ($\mathit{ms}$) and the session keys: ($k_I$) for the client, and ($k_R$) for the server, using Pseudo Random Functions PRFs as follows: (kdf\textsubscript{ms}) takes the $\mathit{pms}$ and nonces as input and produces the $\mathit{ms}$, while (kdf\textsubscript{k}) takes the $\mathit{ms}$ and nonces as input and produces the session keys $k_I$ and $k_R$. There are more than a pair for the session keys, i.e. separate key pairs for encryption and authentication, but we abstract away from these details and refer to the session keys in general by the key pair $k_I$ and $k_R$. Finally, the client sends \texttt{ChangeCipherSpec} (\texttt{CCS}) (this message is not considered part of the handshake and is not included in the transcript hash), followed by a \texttt{ClientFinished} (\texttt{CF}) which is encrypted by the just negotiated algorithms and keys. The \texttt{ClientFinished} verifies the integrity of the handshake transcript (i.e. the $\mathit{log}$ (We adopted the term $\mathit{log}$ from \cite{bhargavan2016downgrade}). The \texttt{ClientFinished} content is computed using a PRF which serves as a Message Authentication Code (MAC) that we denote it by (mac) over a hash of the handshake transcript starting from the \texttt{ClientHello} up to, but not including, the \texttt{ClientFinished} (i.e. mac of $log_1$ as shown in Figure \ref{fig:TLS-ECDHE} and Figure \ref{fig:TLS-RSA}), using the $\mathit{ms}$ as a key. This mac needs to be verified by the server.\par  

Fourth, similar to the client, the server sends its \texttt{ChangeCipherSpec} (\texttt{CCS}) followed by a \texttt{ServerFinished} (\texttt{SF}) that consists of a mac over a hash of the server's transcript up to this point ($log_2$), which also needs to be verified by the client. \par  

Once each communicating party has verified its peer's \texttt{Finished} message, they can now send and receive encrypted data using the established session keys $k_I$ and $k_R$. If \quotes{False Start} \cite{langley2016} is enabled, the client can send data just after its \texttt{ClientFinished}, and before it verifies the \texttt{ServerFinished}.

\subsection{TLS~1.3 Handshake, Major Changes}
\begin{figure}[!tp]
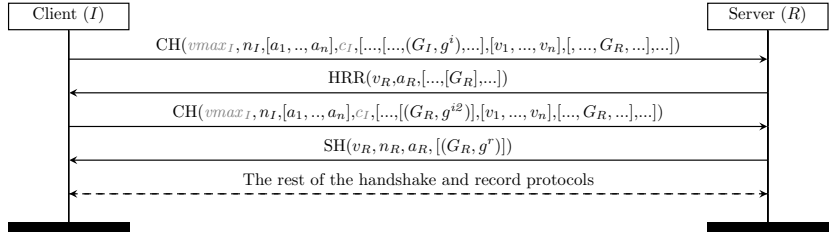
 
\centering
\resizebox{\textwidth}{!}{
\setmsckeyword{} 
\drawframe{no} 

\begin{msc}[normal values, /msc/level height=0.7cm, /msc/label distance=0.1cm , /msc/first level height=0.6cm, /msc/last level height=0.6cm]{} 
\setlength{\instwidth}{2.5\mscunit} 
\setlength{\instdist}{12\mscunit} 
\declinst{I}{}{Client ($I$)}
\declinst{R}{}{Server ($R$)}

\mess{CH($\textcolor{gray}{\mathit{vmax}_{I}},n_I,$[$a_1,..,a_n$],$\textcolor{gray}{c_I}$,[...,[...,($G_I,g^i$),...],[$v_1,...,v_n$],[$,...,G_R,...$],...])} {I}{R}
\nextlevel

\mess{HRR($v_R$,$a_R$,[...,[$G_R$],...])} {R}{I}
\nextlevel

\mess{CH($\textcolor{gray}{\mathit{vmax}_{I}},n_I,$[$a_1,..,a_n$],$\textcolor{gray}{c_I}$,[...,[($G_R,g^{\mathit{i2}}$)],[$v_1,...,v_n$],[$...,G_R,...$],...])} {I}{R}
\nextlevel

\mess{SH($v_R,n_R,a_R,[(G_R,g^r$)])} {R}{I}
\nextlevel

\mess*{The rest of the handshake and record protocols}{I}{R}
\mess*{}{R}{I}

\end{msc}
} 
\caption{Message sequence diagram for TLS~1.3 \texttt{Hello} messages with DHE key-exchange and \texttt{HelloRetryRequest}. Deprecated parameters that are included for backward compatibility are marked with \textcolor{gray}{gray} color.}
\label{fig:TLS13}
\end{figure}

This section is not meant to provide a comprehensive description of TLS~1.3, but to highlight some major changes in TLS~1.3 over its predecessor TLS~1.2. Similar to the previous section, we assume certificate-based unilateral server-authentication mode. A full description of the latest draft of TLS~1.3 (as of this writing) can be found in \cite{tls13rev25}. Figure \ref{fig:TLS13} illustrates the \texttt{Hello} messages in TLS~1.3, where the TLS version and algorithms are negotiated. \par

One of the first changes in TLS~1.3 is prohibiting all known weak and unrecommended cryptographic algorithms such as RC4 for symmetric encryption, RSA and static DH for key-exchange, etc. In addition, TLS~1.3 enforces Forward Secrecy (FS) in both modes: the full handshake mode and the session resumption mode (with the exception of the early data in the Zero Round Trip Time (0-RTT) mode that is always sent in non-FS mode), compared to TLS~1.2, where FS is optional in the full handshake mode, and not possible in the session resumption mode. It also enforces Authenticated Encryption (AE) and standard (i.e. non arbitrary) DH groups and curves. Furthermore, unlike TLS~1.2 where all handshake messages before the \texttt{Finished} messages are sent in cleartext, all TLS~1.3 handshake messages are encrypted as soon as both parties have computed shared keys, i.e. after the \texttt{ServerHello} message. \par

The \texttt{ClientHello} message in TLS~1.3 has major changes. First, in terms of parameters, the following parameters have been deprecated (but still included for backward compatibility): the maximum supported TLS version ($\mathit{vmax}_I$) has been substituted by the \quotes{supported_versions} extension ($[v_1,...,v_n]$); the session ID ($\mathit{session}_{\mathit{ID}}$) has been substituted by the \quotes{pre_shared_key} extension; the compression methods list [$c_1$,...,$c_n$] are not used any more and sent as a single byte set to zero ($c_I$). In addition, unlike TLS~1.2 where extensions are optional, in TLS~1.3, the \texttt{ClientHello} extensions are mandatory and must at least include the \quotes{supported_versions} extension. Second, in terms of behaviour, the server can optionally respond to a \texttt{ClientHello} with a \texttt{HelloRetryRequest} (\texttt{HRR}), a newly introduced message in TLS~1.3 that can be sent from server to client to request a new (EC)DHE group that has not been offered in the client's \quotes{key_share} extension ([...,($G_I,g^i$),...]) which is a list of \quotes{key_share} entries (\quotes{KeyShareEntry}) ordered by preference, but is supported in the client's \quotes{supported_groups} extension ([...,$G_R$,...]). The \texttt{HelloRetryRequest} can also be sent if the client has not sent any \quotes{key_share}. After the \texttt{HelloRetryRequest}, the client sends a second \texttt{ClientHello} with the server's requested \quotes{key_share} ([$G_R,g^{\mathit{i2}}$]). \par 

Upon receiving a \texttt{ClientHello}, if the client's offered parameters are supported by the server, the server responds with a \texttt{ServerHello} message. The \texttt{ServerHello} has two major changes: First, unlike TLS~1.2 where the extensions field is optional, in TLS~1.3, the \texttt{ServerHello} must contain at least the \quotes{key_share} or \quotes{pre_shared_key} extensions (the latter is sent in case of session resumption which is beyond our paper's scope). Second, as a version downgrade attack defence mechanism (in addition to other mechanisms), the last eight bytes of the server's nonce $n_R$ are set to a fixed value that signals the TLS version that the server has received from the client. This allows the client to verify that the versions that were sent in the \texttt{ClientHello} have been received correctly by the server. This is because the nonces are signed in the TLS~1.3 \texttt{CertificateVerify} and in the TLS~1.2 \texttt{ServerKeyExchnage} as well. \par

Finally, the TLS~1.2 \texttt{ServerKeyExchange} is not used in TLS~1.3. This is a result of shifting the key-exchange to the \texttt{Hello} messages, namely to the \quotes{key_share} and \quotes{pre_shared_key} extensions. The signature over the key parameters that is sent in the \texttt{ServerKeyExchange} in TLS~1.2 to authenticate the server's key parameters is now sent in a new message, namely the \texttt{\justify{ServerCertificateVerify}} which is sent after the server's \texttt{Certificate} message. Most importantly, the signature in the \texttt{ServerCertificateVerify} is computed over a hash of the full transcript from the \texttt{Hello} messages up to the \texttt{Certificate}, and not only over the key parameters as in TLS~1.2 \texttt{ServerKeyExchange}. The signature over the full transcript provides protection against downgrade attacks that exploit the lack of ciphersuite authentication in the \texttt{ServerKeyExchange} as demonstrated in \cite{adrian15} and \cite{ beurdouche15}. 
\end{subappendices}
\end{document}